\documentclass[aps,prl,showpacs,superscriptaddress,amsmath,amssymb,twocolumn,preprintnumbers]{revtex4}
\usepackage{graphicx}
\usepackage{dcolumn}
\usepackage{bm}
\usepackage{amsmath}
\usepackage{amssymb}
\usepackage[english]{babel}

\newcommand{\dd}{{\rm d} }
\newcommand{\sqrts}{\sqrt{s}}

\def\pt{p_{_\perp}}
\def\ptaverage{\langle p_{_\perp} \rangle}
\def\xt{x_{_\perp}}
\def\X{{\rm X}}
\def\neff{n_{_{\rm eff}}}
\def\nactive{{n}_{\rm active}}
\def\nspec{{n}_{\rm spectator}}
\def\nexp{n^{{\rm exp}}}
\def\nexpga{n^{{\rm exp}}_{{\gamma}}}
\def\nexpjets{n^{{\rm exp}}_{{\rm jets}}}
\def\nnlo{n^{{\rm NLO}}}
\def\nnloga{n^{{\rm NLO}}_{_{\gamma}}}
\def\thetacm{\vartheta}
\def\sigmainv{\sigma^{\rm inv}}

\def\Deltafit{{\Delta^{\rm fit}}}
\def\sigmamod{\sigma^{\rm model}}

\begin{document}
\preprint{LAPTH-1359/09}
\preprint{SLAC-PUB-13839}

\title{Higher-Twist Dynamics in Large Transverse Momentum Hadron Production}

\author{Fran\c{c}ois Arleo}
\affiliation{Laboratoire d'Annecy-le-Vieux de Physique Th\'eorique (LAPTH), UMR5108, Universit\'e de Savoie, CNRS,\\ BP 110, 74941 Annecy-le-Vieux cedex, France}
\author{Stanley J. Brodsky}
\affiliation{SLAC National Accelerator Laboratory, Stanford University, Stanford,
California 94309, USA}
\author{Dae Sung Hwang}
\affiliation{Department of Physics, Sejong University, Seoul 143--747, Korea}
\author{Anne M. Sickles}
\affiliation{Brookhaven National Laboratory, Upton, New York 11973, USA}

\date{\today}

\begin{abstract}
A scaling law analysis of  the world data on inclusive large-$\pt$ hadron production in hadronic collisions is carried out. A significant deviation {from} leading-twist perturbative QCD predictions at next-to-leading order is reported. The observed discrepancy is largest at high values of $\xt~=~2\pt/\sqrt{s}$. In contrast, the production of prompt photons and jets exhibits the scaling behavior which is close to the conformal limit, in agreement with {the leading-twist expectation}. These results bring evidence for a non-negligible contribution of higher-twist processes in large-$\pt$ hadron production in hadronic collisions, where the hadron is produced directly in the hard subprocess rather than {by} gluon or quark jet fragmentation. Predictions for scaling exponents at RHIC and LHC are given, and it is suggested to trigger the \emph{isolated} large-$\pt$ hadron production to enhance higher-twist processes.
\end{abstract}


\maketitle

\setcounter{footnote}{0}
\renewcommand{\thefootnote}{\arabic{footnote}}

The production of a hadron at large transverse momentum, $\pt$, in a hadronic collision is conventionally analyzed within the framework of perturbative QCD by convoluting the leading-twist (LT) $2 \to 2$ hard subprocess cross sections with evolved structure and fragmentation functions. The most important discriminant of the twist of a perturbative QCD  subprocess in a hard hadronic  collision is the scaling of the inclusive invariant cross section~\cite{Brodsky:1973kr,Tannenbaum0904},
\begin{equation}\label{eq:scaling}
\sigma^{\rm inv} \equiv E\ \frac{\dd\sigma}{\dd^3 p}(A\ B\  \to C\ \X)  = \frac{F(\xt, \thetacm)}{\pt^n},
\end{equation}
at fixed $\xt = {2\pt/ \sqrts}$ and center-of-mass (CM) angle $\thetacm$.  In the original parton model~\cite{Berman:1971xz} the power fall-off is simply  $n=4$ since the underlying $2 \to 2$ subprocess amplitude for point-like partons is scale invariant, and there is no dimensionful parameter as in a conformal theory.
However, in general additional higher-twist (HT) contributions involving a larger number of elementary fields contributing to the hard subprocess, $\nactive>4$,  are also expected.   For example, the detected hadron $C$ can be produced directly in the hard subprocess reaction as in an exclusive reaction.
Such direct HT processes can give a significant contribution since there is no suppression from jet fragmentation at large momentum fraction carried by the hadron, $z$,
and the trigger hadron is produced without any waste of energy. 

Apart from scaling violations due to the QCD running coupling and the evolution of parton distributions functions (PDF) and fragmentation functions (FF), the invariant cross section of a given hard subprocess is expected to scale quite generally as (neglecting spin corrections)~\cite{Brodsky:1994kg}
\begin{equation}\label{eq:scalingrules}
\sigmainv(A\ B\to C\ \X)\propto \frac{(1-\xt)^{2\nspec-1}}{\pt^{2\nactive-4}},
\end{equation}
where $\nspec$ is the number of constituents of $A$, $B$, and $C$ not participating in  the subprocess. From Eq.~(\ref{eq:scalingrules}), it is clear that HT processes involving a large number of active fields will result into a $\pt$-exponent larger than the LT expectation ($n>4$), but will exhibit a slower fall-off with $\xt$ from the smaller number of spectator fields. Therefore, at large $\xt$ and not too large $\pt$, HT contributions to the
cross section can become significant, leading to {an effective exponent higher than the LT expectation}. In Ref.~\cite{Berger:1980qg} the cross sections of the HT subprocesses $g q \to \pi q$ and $q {\bar{q}} \to \pi g$, where the pion is produced directly within the hard subprocess, have been calculated quantitatively in perturbative QCD and compared the results with the cross sections of the LT processes. 
This gives a contribution to ${E\sigma/d^3p}( A B \to \pi \X) $  with nominal scaling $n=6$ at fixed $\xt$ and $\vartheta$ since $\nactive=5$~\cite{Berger:1980qg}.

\begin{figure}[htbp]
  \begin{center}
    \includegraphics[width=8.2cm]{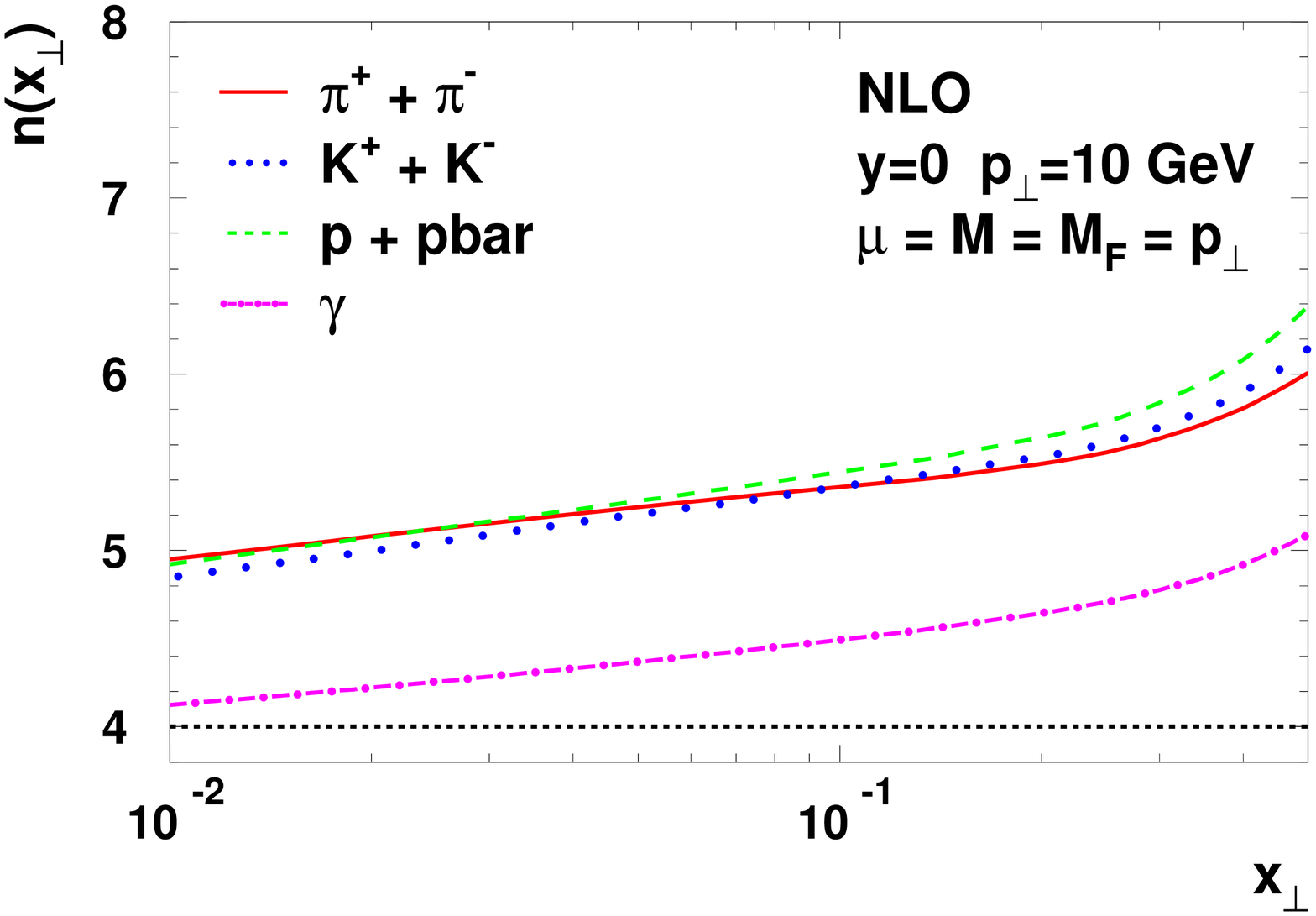}
  \end{center}
\caption{$\xt$-dependence of $\nnlo$ for $\pi^\pm$ (solid line), $K^\pm$ (dotted), $p/\bar{p}$ (dashed) and $\gamma$ (dot-dashed), at $\pt=10$~GeV.}
  \label{fig:xtdep}
\end{figure}

In this Letter, the exponent $\nnlo$ of mid-rapidity particle production ($\vartheta=\pi/2$) 
is computed in QCD at next-to-leading order (NLO) accuracy from~\cite{Aurenche:1998gv}, 
with CTEQ6.6~\cite{Pumplin:2002vw} PDF and the de~Florian--Sassot--Stratmann and Bourhis--Fontannaz--Guillet FF into hadrons and photons~\cite{deFlorian:2007aj}, respectively.
The $\xt$-dependence of $\nnlo$ at fixed $\pt=10$~GeV is shown in Fig.~\ref{fig:xtdep} for pions, kaons, protons/antiprotons, and inclusive prompt photons. The hadron exponents increase slowly from $\nnlo\simeq5$ at small values of $\xt$ ($\xt=10^{-2}$) up to $\nnlo\simeq6$ at $\xt=0.5$; there is very little dependence on {the specific hadron species}. The exponent extracted {in the prompt photon channel is below those of} hadrons, by roughly one unit. The smaller photon exponent is understood from the (relative) absence of fragmentation processes
and one less power in $\alpha_s$, leading to {less scaling violation} in this channel. Remarkably, $\nnloga$ is close to the conformal limit, $n=4$, at the smallest values of $\xt$.

\begin{table*}[htb]
  {\small
  \begin{center}
  \begin{tabular}[c]{p{2.6cm}ccr@{ , }lr@{~--~}lr@{~--~}lc||c|c}
    \hline
    \hline
    Exp. & Ref. &  ~~Species~~ & \multicolumn{2}{c}{$\sqrt{s}$} & \multicolumn{2}{c}{$\pt$} &\multicolumn{2}{c}{~~$x_{_{\perp}}$~~}  & ~~$n_{_{\rm data}}$~~ & $\langle n^{\rm exp}\rangle$ & $\langle \nnlo\rangle$\\
    \hline
    E706 & ~\cite{Apanasevich:2002wt} &  $\pi^0$ &31.6 & 38.8 & 2&9& $10^{-1}$&$4\times10^{-1}$&25 & $ 8.2\pm  0.11$ & $ 6.1\pm  0.09$  \\
PHENIX/ISR & ~\cite{Adare:2008qb,Arleo:2008zd}&  $\pi^0$& 62.4 & 22.4 & 2&7& $2\times10^{-1}$& $2\times10^{-1}$&3& $ 7.5\pm  0.19$& $ 6.2\pm  0.30$  \\
PHENIX &~\cite{Adare:2007dg,Adare:2008qb}&  $\pi^0$& 62.4 & 200 & 2&19 &$7\times10^{-2}$&$2\times10^{-1}$&12& $ 6.7\pm  0.05$ & $ 5.6\pm  0.08$\\
 UA1&	~\cite{Albajar:1989an} &  $h^\pm$ & 500 & 900& 2&9&$8\times10^{-3}$&$2\times10^{-2}$&18 & $ 5.7\pm  0.09$& $ 5.2\pm  0.04$\\
 CDF &~\cite{Abe:1988yu} &  $h^\pm$ &  630 & 1800 & 2&9~&$7\times10^{-3}$&$10^{-2}$&5 & $ 5.2\pm  0.15$ & $ 5.0\pm  0.07$\\
 CDF &~\cite{Acosta:2001rm} &  tracks &  630 & 1800 & 2&19&$7\times10^{-3}$&$2\times10^{-2}$& 52& $ 5.7\pm 0.03 $ & $ 5.0\pm  0.02$\\
\hline
 CDF&~\cite{Acosta:2002ya} &  $\gamma$&  630 & 1800& 11&81&$3\times10^{-2}$&$9\times10^{-2}$& 7 & $ 4.7\pm  0.09$& $ 4.3\pm  0.01$ \\
 D0& ~\cite{Abbott:1999kd,Abazov:2001af}&  $\gamma$& 630 & 1800 & 11&107&$3\times10^{-2}$&$10^{-1}$& 6 & $ 4.5\pm  0.12$& $ 4.3\pm  0.01$ \\
\hline
 CDF & ~\cite{Abe:1992bk}& jets&  546 & 1800&29&190 &$10^{-1}$&$2\times10^{-1}$&9 & $ 4.3\pm  0.09$& $ 4.6\pm  0.01$\\
 D0&~\cite{Abbott:2000kp} & jets &  630 & 1800& 23&376 &$8\times10^{-2}$&$4\times10^{-1}$& 23& $ 4.5\pm  0.04$& $ 4.6\pm  0.01$\\
    \hline
    \hline
\end{tabular}
  \caption{Data sets selected in the present Letter. The kinematical range ($\sqrt{s}$, $\pt$ in GeV), the mean $\langle\nexp\rangle$ extracted from each set composed of $n_{_{\rm data}}$ data points and the corresponding expectation in QCD at NLO, $\langle\nnlo\rangle$, are given.}
  \label{tab:data}
  \end{center}}
\end{table*}

In order to investigate possible HT dynamics in large-$\pt$ hadron production, $\nexp$ has been systematically extracted from measurements in $p$--$p$ and $p$--$\bar{p}$ collisions, from fixed-target to collider experiments, and compared to LT QCD expectations. It is deduced from the comparison of $\xt$-spectra at different CM energies,
\begin{equation}\label{eq:nexp}
\nexp(\xt) \equiv-\frac{\ln\left(\sigmainv(\xt,\sqrt{s_1})\big/\sigmainv(\xt,\sqrt{s_2}) \right)}{\ln\left( \sqrt{s_1}\big/\sqrt{s_2} \right)}   
\end{equation}
which is equivalent to (\ref{eq:scaling}) at fixed $\xt$. In order to reduce systematic uncertainties, only experiments which measured $\xt$-spectra at two distinct CM energies are considered, except for the PHENIX results at $\sqrt{s}=62.4$~GeV~\cite{Adare:2008qb} compared to a fit of ISR measurements at $\sqrt{s}=22.4$~GeV~\cite{Arleo:2008zd}. The recent data analyzed in this Letter are summarized in Table~\ref{tab:data}. The data sets include $\pi^0$ measurements by the E706 at FNAL~\cite{Apanasevich:2002wt} 
and by the PHENIX collaboration at RHIC~\cite{Adare:2007dg,Adare:2008qb}.
At higher energies, the measurements of charged hadrons (or charged tracks~\cite{Acosta:2001rm}) in $p$--$\bar{p}$ collisions at 
$\sqrts=630,\ 1800$~GeV by CDF~\cite{Abe:1988yu,Acosta:2001rm}  
and $\sqrts=500,\ 900$~GeV by UA1~\cite{Albajar:1989an}  
are included in the analysis. Also considered are prompt photon~\cite{Acosta:2002ya,Abbott:1999kd,Abazov:2001af} and jet~\cite{Abe:1992bk,Abbott:2000kp} data obtained 
by CDF and D0 at $\sqrts=546,\ 630,\ 1800$~GeV.

\begin{figure*}[t]
  \begin{center}
\begin{minipage}{7.cm}
    \includegraphics[height=6.5cm]{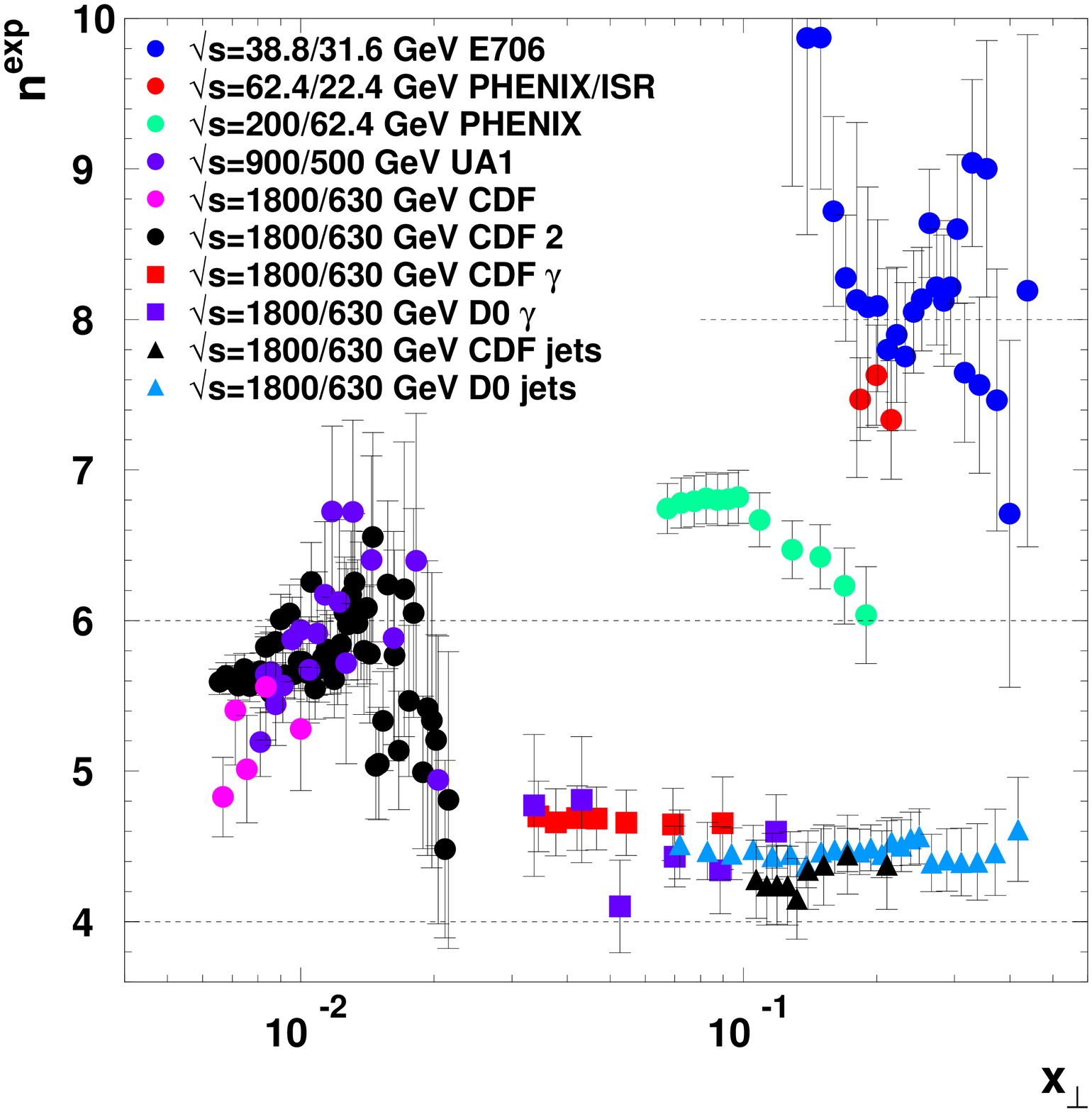}
\end{minipage}
~
\begin{minipage}{10.cm}
    \includegraphics[width=10.cm,height=6.5cm]{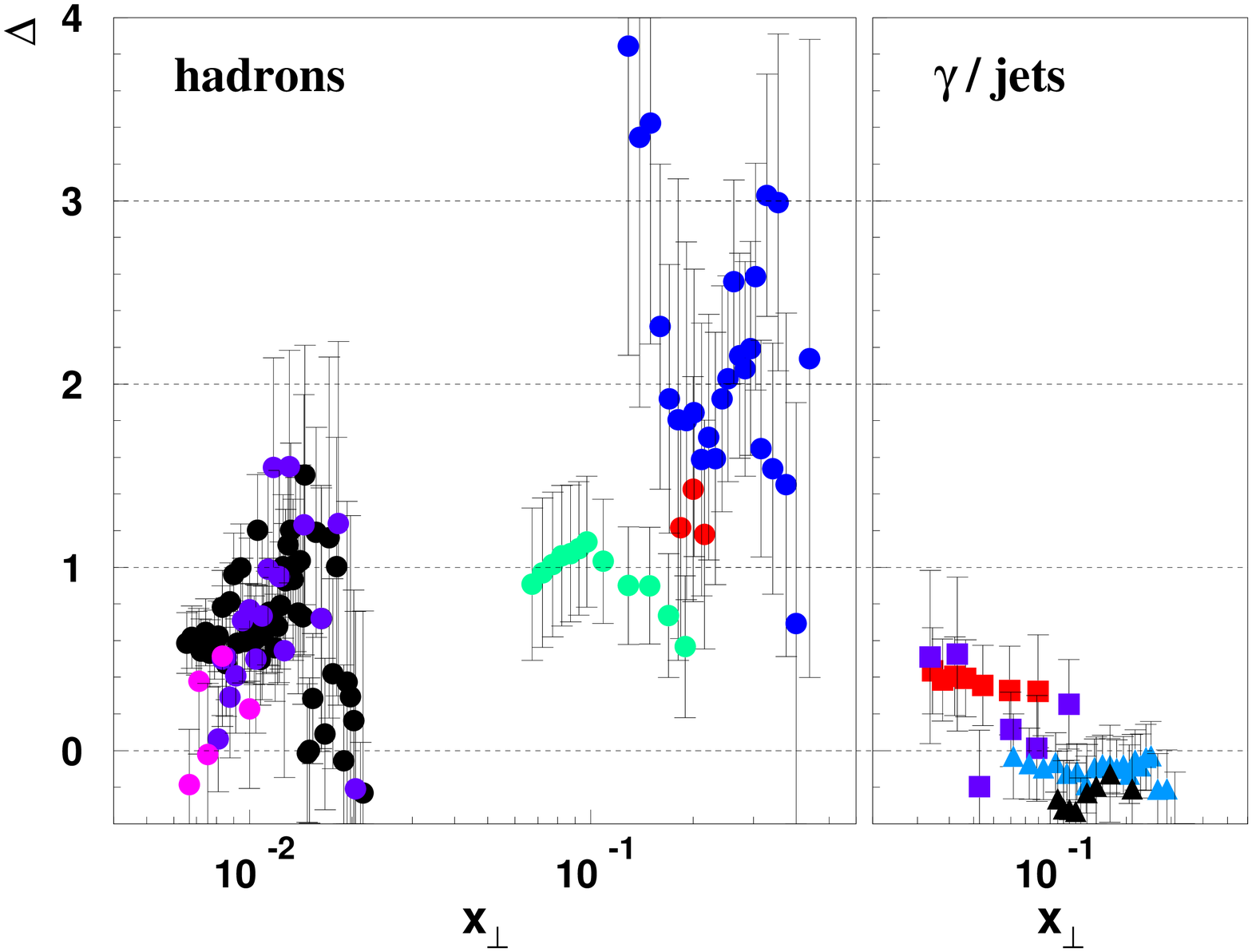}
\end{minipage}

\caption{{\it Left:} Values of $\nexp$ as a function of $\xt$ for $h^\pm/\pi^0$ (circles), $\gamma$ (squares) and jets (triangles). {\it Right: }$\Delta\equiv\nexp-\nnlo$ as a function of $\xt$, error bars include the experimental and the theoretical uncertainties added in quadrature (see text).}
  \label{fig:datatheory_compilation}
  \end{center}
\end{figure*}

The hadron exponents plotted in Fig.~\ref{fig:datatheory_compilation} (left) exhibit a clear trend, with a significant rise of $\nexp$ as a function of $\xt$. Typical values of $\nexp$ are $\nexp\simeq5$--$6$ at small $\xt\simeq10^{-2}$ while PHENIX data 
point to a mean value $\nexp=6.7\pm 0.05$ at $\xt\simeq10^{-1}$. At higher values of $\xt$, the comparison of PHENIX with ISR data as well as the E706 measurements reveal an exponent even larger: $\nexp=7.5\pm0.19$ ($\xt=0.2$) and $\nexp=8.2\pm0.11$ ($\xt=0.2$--$0.4$), respectively. The E706 data clearly confirm results reported long ago at the ISR, which are re-analyzed in a forthcoming paper~\cite{ABHW:2009}.
The results obtained in the photon and jet channels are strikingly different from what is observed for hadrons. Their exponents show almost no dependence on $\xt$, yet the data cover a wide complementary range: $\xt=0.04$--$0.1$ for photons and  $\xt=0.08$--$0.4$ for jets. Importantly enough, the values obtained lie only slightly above the conformal limit, $\nexpga\simeq4.6$ and $\nexpjets\simeq4.4$; most significantly they are  several units smaller than the hadron exponents taken at the same $\xt$ (the $\pt$ range being however different).

In order to compare properly data and theory, NLO calculations have also been carried out within the same kinematical conditions as the experiments. The \emph{difference} between experimental and theoretical exponents, $\Delta(\xt)\equiv\nexp-\nnlo$, is plotted in the right panel of Fig.~\ref{fig:datatheory_compilation} for hadrons and photons/jets.  Note that the error bars include both experimental \emph{as well as} theoretical errors, added in quadrature. The biggest theoretical uncertainty comes from the variation of renormalization/factorization scales, for which all scales were varied from $\pt /2$ to $2\pt$, as is common practice (the renormalization scale ambiguity can be removed using the methods described in~\cite{Binger:2006sj}).
Fig.~\ref{fig:datatheory_compilation} (right) indicates that the hadronic exponents extracted experimentally prove significantly above the LT predictions.
The discrepancy is moderate at small $\xt$, $\Delta(\xt\sim10^{-2})\simeq0.5$, but becomes increasingly larger at higher values of $\xt$: the PHENIX measurements at $\xt\simeq10^{-1}$
lead to $\Delta\simeq1$ and the exponent inferred from E706 data
 is two units above LT expectations.  In contrast, the scaling behaviors observed for photons and jets
remarkably coincide, in excellent agreement with the NLO predictions.
Part of the discrepancy between data and fixed-order calculations at large $\xt\sim1$ could occur  because of the appearance of large threshold logarithms, $\ln(1-\xt)$, which should be resummed to all orders \cite{Laenen:2000ij}. However, the discrepancy is also observed at small values of $\xt\sim10^{-2}$, where threshold effects are expected to be small.

The most natural explanation for the hadron data is the presence of important HT contributions from processes in which the detected hadron appears in the hard subprocess. The dimension of the hadron distribution amplitude  leads naturally larger exponents; see Eq.~(\ref{eq:scalingrules}).  In contrast, particles having no hadronic structure like isolated photons and jets are much less sensitive to such HT contributions and should behave closer to LT expectations, as observed. Another piece of evidence for HT effects is the larger exponents for protons than for pions observed at the ISR~\cite{ABHW:2009}. According to Eq.~(\ref{eq:scalingrules}), the exponent {of HT} would be $n_{\pi}=6$ for pions ($\nactive=5$) and $n_p=8$ for protons ($\nactive=6$), leading to $n_{p}-n_{\pi}=2$ instead of $n_{p}-n_{\pi}\simeq0$ at LT
(see Fig.~\ref{fig:xtdep}). 
  The experimental value obtained from the ISR~\cite{ABHW:2009},
 $n_{p}-n_{\pi}\simeq1$, thus reflects the mixture of LT and HT contributions to the total cross section.   It has been noted~\cite{Brodsky:2008qp} that the presence of color-transparent  HT subprocesses such as $ u u \to p \bar d$ can account for the anomalous features of proton production seen in heavy ion collisions at RHIC~\cite{Adler:2003kg}.

\begin{figure}[htbp]
\begin{center}
\includegraphics[height=6.1cm]{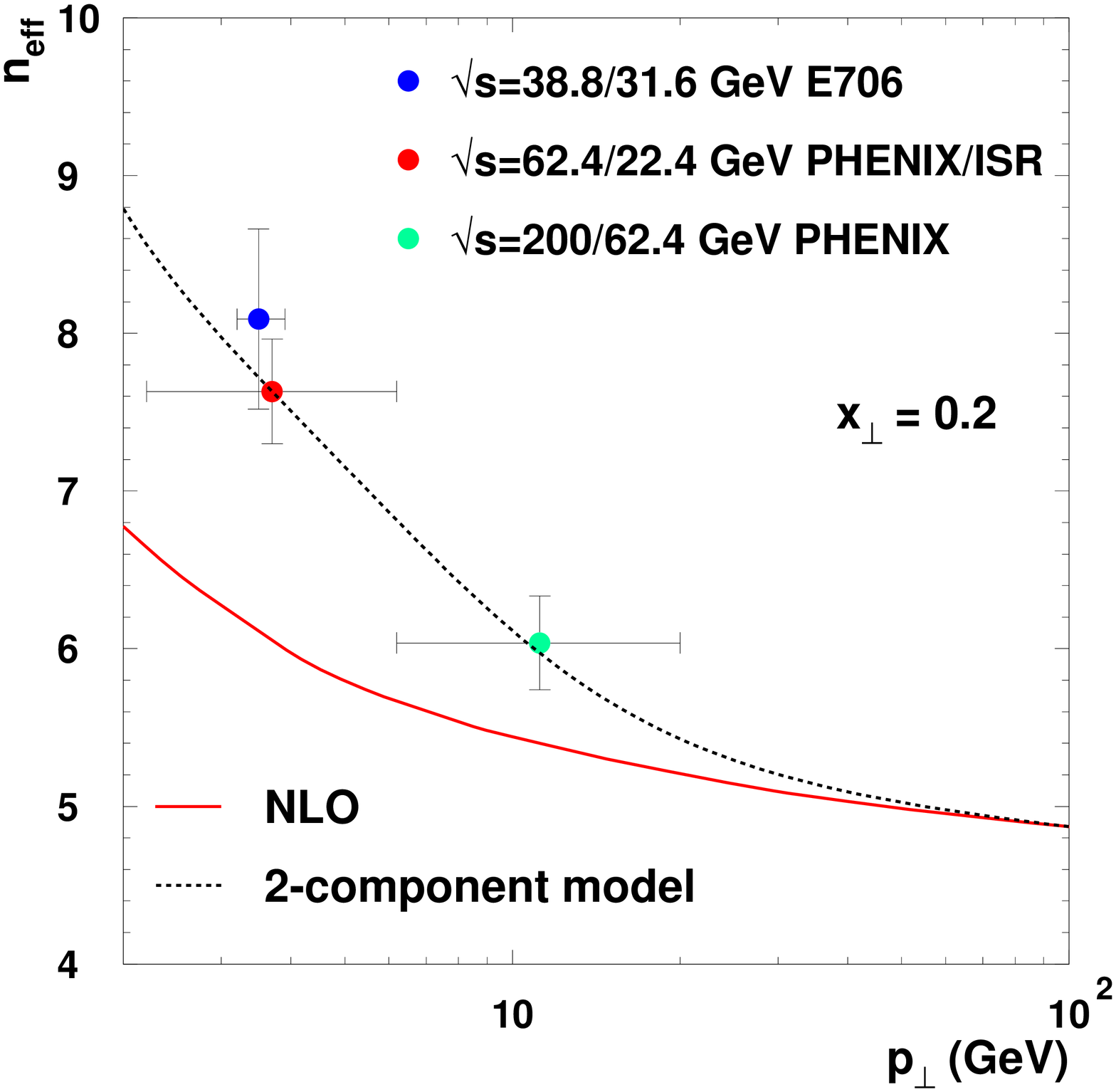}
\end{center}
\caption{$\pt$-dependence of $\neff$ of pions at $\xt=0.2$ in QCD at NLO (solid line). The dotted line represents a fit based on a two-component model with $B/A=50$~GeV$^2$, see Eq.~(\ref{eq:neff2}).}
\label{fig:scaledep}
\end{figure}

In order to probe the HT effect explicitly, let us consider a 2-component model cross section with nominal power dependence
\begin{equation}\label{eq:scalingrulespi}
\sigmamod(p p \to \pi\ \X)\propto \frac{A(\xt)}{\pt^4} + \frac{B(\xt)}{\pt^6},
\end{equation}
corresponding to the LT ($\nactive=4$) and HT ($\nactive=5$) processes, respectively.
The actual $\pt$-exponents are modified by the running coupling and PDF and FF evolution.
Assuming that the contributions  to $\nnlo-4 $ due to pQCD  are the same for the LT and HT processes, Eq. (\ref{eq:scalingrulespi})
gives the \emph{effective} exponent
\begin{eqnarray}\label{eq:neff2}
\neff(\xt,\pt,B/A) &\equiv& -\frac{\partial\ln \sigmamod}{\partial\ln\pt}+\nnlo(\xt,\pt)-4\nonumber\\
&=& \frac{2B/A}{\pt^2+B/A}+\nnlo(\xt,\pt).
\end{eqnarray}
Note that $\neff  \to \nnlo+ 2$ for $B/A \to \infty.$
As shown in Fig.~\ref{fig:scaledep} (solid line), the LT pion exponent (evaluated at $\xt=0.2$) slowly decreases
with $\pt$ and reaches $n=4$ as $\pt\to\infty$ because of asymptotic freedom.
Eq. (\ref{eq:neff2}) shows that $\neff$ depends on the relative
strength of HT corrections to the LT cross section, $B/A$.
The value $B/A\sim50$~GeV$^2$ is extracted from the data as shown by the dotted line in Fig.~\ref{fig:scaledep}.
However, a somewhat smaller estimate, $B/A\sim15$~GeV$^2$, is obtained
when all scales are set to $\pt/2$ in the QCD calculation.
We note that the HT rate for direct processes and therefore $B/A$ are
enhanced relative to fragmentation processes since the trigger hadron is
produced without any waste of energy;  thus the magnitude of the subprocess amplitude is maximized since it is evaluated at the trigger $\pt$, and the initial momentum  fractions $x_1$ and $x_2$ are evaluated at small values where the PDF are largest.

\begin{figure}[htbp]
\begin{center}
\includegraphics[height=4.8cm]{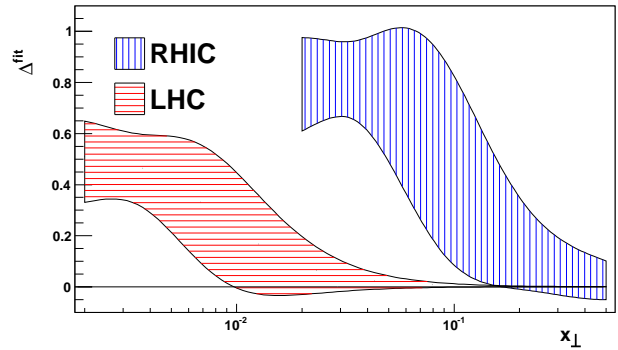}
\end{center}
\caption{Predicted difference between the experimental and NLO scaling 
exponent at RHIC $\sqrt{s}=200,500 $ GeV and the LHC  ($\sqrt{s}=7$~TeV as compared to $\sqrt{s}=1.8$~TeV) based on a global fit of existing RHIC and Tevatron data }
\label{fig:lhc}
\end{figure}

Finally, we discuss the phenomenological consequences of possible HT contributions to hadron production in $p$--$p$ collisions at RHIC and LHC. In order to obtain qualitative predictions, the difference $\Delta$ between the experimental and the NLO exponent has been fitted to the hadron data in Table~\ref{tab:data} using a simple parametrization (with $\ptaverage$ the geometrical mean of the two experimental $\pt$-bins)
\begin{equation}\label{eq:fit}
\Deltafit({\xt}, {\ptaverage})= p_0\left( -\log {\xt} \right)^{p_1}  \times  \frac{2\ p_2 (1-\xt)^{p_3}}{{\ptaverage}^2 + p_2 (1-\xt)^{p_3}},\nonumber
\end{equation}
inspired by the 2-component model above described. As expected in QCD, $\Deltafit$ is vanishing in $\pt\to\infty$ limit at fixed $\xt$.
This analytic form 
is somewhat arbitrary but flexible enough for making predictions beyond the $(\xt,\pt)$-range probed in present experiments. The typical values of $\Deltafit$ expected at RHIC (taking $\sqrts=200, 500$~GeV) and at the LHC  ($\sqrts=7$~TeV, compared to $\sqrts=1.8$~TeV at Tevatron) are plotted as a function of $\xt$ in Fig.~\ref{fig:lhc}. At RHIC, $\Deltafit$ is slightly below 1 at small $\xt\lesssim5.10^{-2}$ but decreases towards zero at larger $\xt$ (i.e. larger $\pt$). At LHC, smaller deviations with NLO expectations are expected because of the large values of $\ptaverage$ probed at high energy: $\Deltafit\simeq0.5$  below $\xt=5\times10^{-3}$ (corresponding to $\pt\sim20$~GeV at $\sqrt{s}=7$~TeV) and smaller above.
From this, the ratios of $\xt$-spectra can be determined straightforwardly, $R_{\sqrt{s_1}/\sqrt{s_2}}=(\sqrt{s_2}/\sqrt{s_1})^{\Deltafit+\nnlo}$, where the NLO exponents at RHIC ($\nnlo\simeq5.3$) and LHC ($\nnlo\simeq4.8$) do not vary significantly in the considered $\xt$ range. 
In order to enhance the HT contribution to hadron production, we suggest to trigger on {\it isolated} hadrons, i.e. with small hadronic background in their vicinity. The use of isolation cuts, usually applied for prompt photons, will strongly suppress LT processes. As a consequence, the scaling exponents of isolated hadrons are expected to be somewhat larger than in the inclusive channel.

The evidence for higher-twist dynamics reported in this analysis supports the interpretation of heavy-ion collision measurements at RHIC, in which the dense QCD medium enhances HT  contributions, and thus proton production, by filtering LT processes due to partonic energy loss~\cite{Brodsky:2008qp}. 
Future RHIC and LHC measurements will provide further tests of the dynamics of large-$\pt$ hadron production beyond leading twist.

FA thanks P. Aurenche for useful discussions and CERN-TH for hospitality. SJB was supported by the Department of Energy under contract DE-AC02-76SF00515.  DSH was supported by the International Cooperation 
Program of the KICOS and the Korea Research Foundation Grant (KRF-2008-313-C00166).  AMS was supported by the Department of Energy
under contract DE-AC02-98CH10886.

\providecommand{\href}[2]{#2}\begingroup\raggedright\endgroup

\end{document}